\documentclass[10pt,final,conference,letterpaper,romanappendices,twocolumn]{IEEEtran}
\usepackage{cite}
\usepackage{textcomp}
\usepackage{xcolor}
\usepackage{epstopdf}
\usepackage{epsfig}
\usepackage{amsthm}
\usepackage{amsmath,amssymb,amsfonts,amstext,amsbsy,amsopn,dsfont}
\usepackage{enumitem}
\usepackage{cases}
\usepackage{sublabel}

\newtheorem{proposition}{\textbf{Proposition}}

\newcommand{\defn}{\triangleq}
\newcommand{\dif}{\mathrm{d}}

\begin{document}

\title{ A 3D Modeling Approach to Tractable Analysis in UAV-Enabled Cellular Networks\\
}

\author{\IEEEauthorblockN{Chun-Hung Liu}
\IEEEauthorblockA{Department of Electrical \& Computer Engineering \\
Mississippi State University, MS, USA \\
e-mail: chliu@ece.msstate.edu}
\and
\IEEEauthorblockN{Di-Chun Liang and Rung-Hung Gau}
\IEEEauthorblockA{Department of Electrical \& Computer Engineering \\
National Chiao Tung University, Hsinchu, Taiwan \\
e-mail: runghunggau@g2.nctu.edu.tw}
}

\maketitle

\begin{abstract}
This paper aims to propose a three-dimensional (3D) point process that can be employed to generally deploy unmanned aerial vehicles (UAVs) in a large-scale cellular network and tractably analyze the fundamental network-wide performances of the network. This 3D point process is devised based on a 2D marked Poisson point process in which each point and its random mark uniquely correspond to the projection and the altitude of each point in the 3D point process, respectively. We elaborate on some important statistical properties of the proposed 3D point process and then use them to tractably analyze the coverage performances of a UAV-enabled cellular network wherein all the UAVs equipped with multiple antennas are served as aerial base stations. The downlink coverage of the UAV-enabled cellular network is found and its closed-form results for some special cases are explicitly derived as well. Furthermore, the fundamental limits achieved by cell-free massive antenna array are characterized when coordinating all the UAVs to jointly perform non-coherent downlink transmission. These findings are validated by numerical simulation.
\end{abstract}

\section{Introduction}
Unmanned aerial vehicles (UAVs) have found a wide range of applications thanks to their outstanding capability of agilely moving in three-dimension (3D) space, which relieves spatial limitations caused by two static terminals. Despite the agile and flexible mobility of UAVs, it may not really facilitate communications in a wireless network where many UAVs are arbitrarily deployed and a considerable amount of co-channel interference is created accordingly. The 3D deploying problem for a UAV-enabled cellular network with UAVs serving as aerial base stations is involved in the issue of simultaneous multi-user coverage, and thereby it is much more complicated and difficult than the 3D deploying problem for a cellular-connected UAV network that merely needs to tackle the issue of single UAV coverage at a time. Deploying methods for a UAV-enabled cellular network should be able to exploit the mobility of UAVs in order to ameliorate the fundamental coverage limit of the entire cellular network, yet how to evaluate the deploying methods in a \textit{tractable} and \textit{network-wide} way remains unclear until now.

There are indeed some prior works that modeled UAV-enabled wireless networks in a large-scale sense, e.g., \cite{VVCHSD17,HTLYSZSXCY1902,SYDZDX19}. However, the majority of them simply assumed that all UAVs in a network hover at the same \textit{fixed altitude}. For example, reference \cite{VVCHSD17} investigated the coverage problem for a finite network model assuming a number of UAVs are uniformly distributed at the same fixed altitude in the network. The coverages based on UAV-centric and user-centric strategies for multi-UAV-assisted NOMA networks were studied in~\cite{HTLYSZSXCY1902}. Reference~\cite{SYDZDX19} proposed a UAV-assisted wireless network for the malfunction areas and used a user-centric cooperation scheme to evaluate the coverage and normalized spectral efficiency of the network. These prior works all assumed that all UAVs hover at the same fixed altitude in a network so that their analyses cannot practically reflect how they are influenced by a real-world deployment of UAVs with a \textit{random} altitude. Some prior works already tried to relax the modeling assumption of ``fixed altitude'' when modeling multiple UAVs in the sky. Reference \cite{SBLLRRMMVRJH20}, for example, studied the coverage probability in a 3D deployment model of UAVs wherein all UAVs were distributed within a specific range of altitude that was uniformly divided into a certain number of levels.

 A few prior works also adopted 3D homogeneous PPPs to model UAV-enabled cellular networks.  Reference~\cite{LCHHKHWJ19} exploited the limits of the coverage and volume spectral efficiency of a mmWave UAV cellular network in which a UAV's altitude was modeled as a function of the UAV's projection. The coverage and network throughput of a NOMA-assisted UAV network modeled by a 3D homogeneous PPP were analyzed in~\cite{HTLYSZSXCY2003}, whereas reference~\cite{ZCZW16} considered spectrum sharing when analyzing the success probability and total network throughput of a UAV-enabled network modeled by a 3D PPP. Modeling the distribution of UAVs by 3D PPPs leads to two practical issues. One is that UAVs are low-altitude platforms and cannot be arbitrarily positioned in infinitely large 3D space modeled by a 3D PPP. The other is that the path-loss exponent of any wireless links in a wireless network modeled by a 3D PPP needs to be greater than three in order to make analysis bounded, yet such a constraint on the path-loss exponent is not practically true for most 3D wireless links with a path-loss exponent smaller than three.

Although these aforementioned prior works successfully conducted some analyses for specific problems, in general their outcomes are not easily generalized to a network-wide scenario in a large-scale UAV-enabled cellular network in that their generality is subject to their simplified models and assumptions of deploying UAVs in a wireless network. In contrast, this paper proposes a 3D modeling approach to deploying large-scale UAV-enabled cellular networks, which is much more general and practical for UAV deployment than the prior works in the literature. As such, we are able to obtain much more accurate analytical results by employing it. Other main contributions of this paper are summarized as follows. By employing the proposed 3D point process to position all the UAVs in the sky, we consider the \textit{angle-projection-independent locating} (APIL) scenario in which the elevation angle and the projection of each UAV are independent, which leads to high tractability in analysis. We  are thus able to explicitly derive the downlink coverage and the \textit{cell-free downlink coverages} for all the UAVs doing non-coherent joint transmission. Furthermore, the numerical results show that in general the downlink coverages are insensitive to the different distributions of the elevation angle and of UAVs that have the same mean so that they can be approximated by the derived expressions using the mean of the elevation angle of a UAV.

\section{The Proposed 3D Point Process Model}
Suppose a 2D homogeneous PPP of density $\lambda$ can be denoted by the following set on the plane of $\mathbb{R}^2$:
\begin{align}
	\Phi_x\defn \{X_i\in\mathbb{R}^2: i\in\mathbb{N}_+\}.
\end{align}
In accordance with $\Phi_x$, we propose the following 3D point process $\Phi_u$:
\begin{align}\label{Eqn:3DPP}
\Phi_u \defn \bigg\{ & U_i\in\mathbb{R}^2\times \mathbb{R}_+: U_i=(X_i,H_i), X_i\in\Phi_x, \nonumber\\
&H_i=\|X_i\|\tan(\Theta_i), \Theta_i\in\left[0,\frac{\pi}{2}\right], i\in\mathbb{N}_+\bigg\},
\end{align}
where $X_i$ is the projection of point $U_i$ on the plane of $\mathbb{R}^2$, $\|X_i\|$ is the distance between the origin\footnote{Without loss of generality, in this paper we use the origin as a reference point for the locations of the points in point sets such as $\Phi_x$ and $\Phi_u$ to express their relevant equations, results, and observations. According to the Slivnyak theorem~\cite{DSWKJM13}\cite{MH12}, the statistical properties of a PPP evaluated at the origin are the same as those evaluated at any particular point in the PPP.} and $X_i$,  and $\Theta_i$ is the (random) elevation angle from the origin to point $U_i$. Hence, the ``altitude'' of point $U_i$ is $H_i$ that is the distance from $X_i$ to $U_i$ such that $\Phi_u$ can be viewed as a marked version of $\Phi_x$ in which each point has a mark as its altitude. Since $\|Y_i-Y_j\|$ denotes the Euclidean distance between points $Y_i$ and $Y_j$ for $i\neq j$, we know $\|X_i\|=\|U_i\|\cos(\Theta_i)$ and thus $\|U_i\|=\|X_i\|\sec(\Theta_i)$. 

\subsection{The LoS Probability of Channels and The APIL Scenario}
A link between two spatial points is called a LoS link provided it is not visually blocked from one point to the other. A low-altitude-platform communication scenario is considered in this paper and the LoS model of a 3D channel in~\cite{AHKSSL14} is adopted so that we have the following LoS probability of the 3D channel between the origin and a point $U_i\in\Phi_u$ proposed in~\cite{AHKSSL14}:
\begin{align}\label{Eqn:LoSProb}
	\rho\left(\Theta_i\right) \defn \frac{1}{1+c_2\exp\left(-c_1\Theta_i\right)},
\end{align}
where $c_1$ and $c_2$ are environment-related positive constants (for rural, urban, etc.), and thereby whether or not point $U_i$ is LoS for the origin is completely determined by its elevation angle $\Theta_i$ from the origin.

\begin{figure}[t!]
	\centering
	\includegraphics[width=3.25in, height=2.15in]{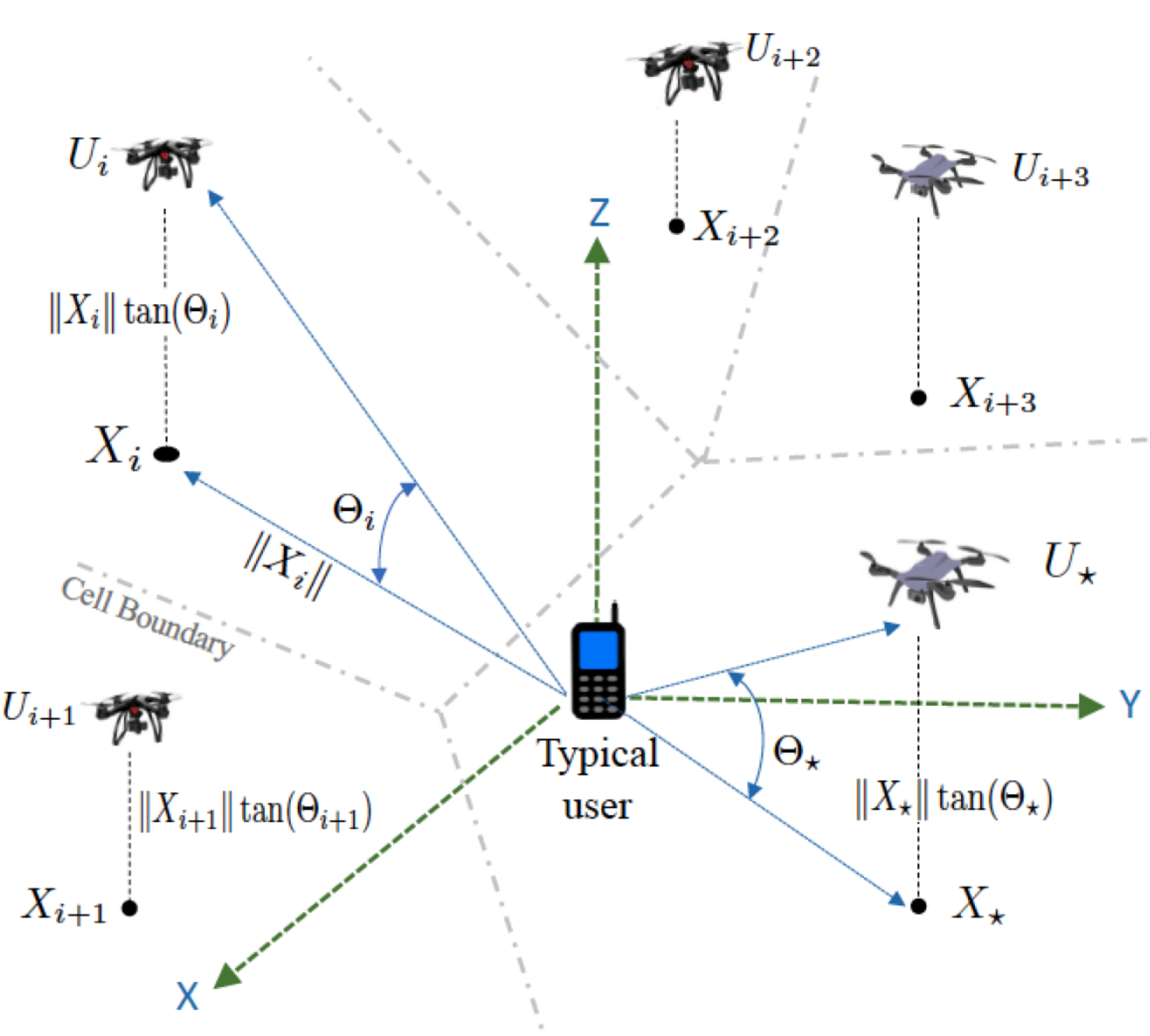}
	\caption{The proposed 3D point process $\Phi_u$ is used to model the locations of the UAVs in a cellular network.  The projection of point $U_i\in\Phi_u$ is denoted by $X_i$ and a typical user located at the origin associates with UAV $U_{\star}$ serving as its aerial base station. The APIL scenario is considered in this paper, that is, $\Theta_i$ and $\|X_i\|$ are independent for all $i\in\mathbb{N}_+$ and $H_i=\tan(\Theta_i)\|X_i\|$ depends on $\Theta_i$ and $X_i$.}
	\label{Fig:SystemModel}
\end{figure}

For the 3D point process $\Phi_u$,  we will specifically consider a positioning scenario in which the elevation angle and projection of each point in $\Phi_u$ are independent, which is referred to as the angle-projection-independent locating (APIL) scenario in this paper. An illustration of the proposed 3D point process $\Phi_u$ for the APIL scenario is depicted in Fig.~\ref{Fig:SystemModel}. In the figure, we employ $\Phi_u$ to model the 3D locations of UAVs in a cellular network and the projections of the UAVs on the $\mathsf{X}-\mathsf{Y}$ (ground) plane form a 2D homogeneous PPP $\Phi_x$. In the APIL scenario, the elevation angle and the projection of each point in $\Phi_u$ are independent, i.e., $\Theta_i$ and $X_i$ are independent for all $i\in\mathbb{N}_+$ and all $\Theta_i$'s are assumed to be identically and independently distributed (i.i.d.) random variables (RVs) in this paper. The APIL scenario properly characterizes the situation that locating point $U_i\in\Phi_u$ is accomplished by using two independent parameters $\Theta_i$ and $X_i$ and such a situation refers to when each point (UAV) in $\Phi_u$ is randomly positioned at a certain elevation angle whose distribution can be observed at the origin. In the following, we will analyze some important statistical properties related to $\Phi_u$ in the APIL scenario.

\subsection{Distance-Related Distributions in $\Phi_u$}\label{SubSec:Dis-RelDis}
Suppose a non-negative RV $R_{\star}$ is defined as
\begin{align}\label{Eqn:Weight3DDis}
	R_{\star}\defn \max_{i:U_i\in\Phi_u} \left\{W_iL_i\|U_i\|^{-\alpha}\right\},
\end{align}
where $\alpha>2$ is a constant\footnote{ If $\|U_i\|^{-\alpha}$ stands for the path loss between node $U_i$ and the origin, $\alpha$ is referred to as the path-loss exponent, which will be used in Section~\ref{Sec:UAVNetworkModel}.}, $L_i\in\{1,\ell\}$ is a Bernoulli RV that is equal to one if an LoS link between the origin and point $U_i$ exists and $\ell$ otherwise, and $W_i\in\mathbb{R}_+$ is a non-negative weighting RV associating with $U_i$ and independent of all $L_i$'s and $U_i$'s. Note that $\ell\in[0,1]$ is referred to as the NLoS channel attenuation factor since it is used to model the penetration loss of an NLoS link, $W_i$ is independent of $L_j$ and $U_j$ for all $i,j\in\mathbb{N}_+$, all $W_i$'s are assumed to be i.i.d., and the distribution of $L_i$ depends on the location of $U_i$ as indicated by the LoS probability in~\eqref{Eqn:LoSProb}. We then have the following theorem.
\begin{proposition}\label{Thm:CDFR*}
	Suppose the moment of $W_i$ exists (i.e., $\mathbb{E}[W_i^{a}]<\infty$ for all $a>0$) for all $i\in\mathbb{N}_+$. If the APIL scenario is considered, the cumulative density function (CDF) of $R_{\star}$ defined in~\eqref{Eqn:Weight3DDis}  can be found as
	\begin{align}\label{Eqn:CDFR*1}
		F_{R_{\star}}(r) = &\exp\left(-\pi\lambda \mathbb{E}\left[W^{\frac{2}{\alpha}}\right]\omega  r^{-\frac{2}{\alpha}}\right),
	\end{align}
	where $F_Z(\cdot)$ denotes the CDF of RV $Z$ and $\omega$ is defined as
	\begin{align}\label{Eqn:Density3DPPInd}
		\omega\defn  \mathbb{E}\left\{\cos^2(\Theta)\left[\rho(\Theta)\left(1-\ell^{\frac{2}{\alpha}}\right) +\ell^{\frac{2}{\alpha}}\right]\right\}.
	\end{align}
\end{proposition}
\begin{IEEEproof}
	See Appendix.
\end{IEEEproof}

The results in Proposition~\ref{Thm:CDFR*} are very general since they are valid for the general distributions of $W$ and $\Theta$. Accordingly, they can be employed to find the distributions of some specific RVs related to set $\Phi_u$. To demonstrate this, we discuss some special cases of $R_{\star}$ in the following.
\subsubsection{$W_i=L_i=1$} In this case, $R_{\star}$ in~\eqref{Eqn:Weight3DDis} reduces to $R_{\star}=\max_{U_i\in\Phi_u}\|U_i\|^{-\alpha}$ so that $R^{-\frac{1}{\alpha}}_{\star}=\min_{U_i\in\Phi_u}\|U_i\|$ is the shortest distance between the origin and set $\Phi_u$. Thus, using $F_{R_{\star}}(r)$ in~\eqref{Eqn:CDFR*1} helps find the CCDF of $R^{-\frac{2}{\alpha}}_{\star}$ as
\begin{align}\label{Eqn:CCDFInvR*1}
	F^c_{R^{-\frac{2}{\alpha}}_{\star}}(y) = \exp\bigg(-\pi\lambda\mathbb{E}\left[\cos^2(\Theta)\right]  y\bigg),
\end{align}
which indicates that $R^{-2/\alpha}_{\star}\sim\exp(\pi\lambda\mathbb{E}[\cos^2(\Theta)])$ is an exponential RV with mean $1/\pi\lambda\mathbb{E}[\cos^2(\Theta)]$, and it is exactly the CCDF of the square of the shortest distance between the origin and  a 2D homogeneous PPP of density $\lambda \mathbb{E}[\cos^2(\Theta)]$~\cite{DSWKJM13,MH12}. Namely, this observation manifests that \textit{the 3D point process $\Phi_u$ proposed in~\eqref{Eqn:3DPP} can be equivalently viewed as a 2D homogeneous PPP of density $\lambda \mathbb{E}[\cos^2(\Theta)]$ as long as the elevation angle and the projection of each point in $\Phi_u$ are independent}.

\subsubsection{$W_i=1$} For this case, $R_{\star}$ in~\eqref{Eqn:Weight3DDis} reduces to $R_{\star}=\max_{i:U_i\in\Phi_u} L_i\|U_i\|^{-\alpha}$ and thus $R^{-\frac{1}{\alpha}}_{\star}=\min_{i:U_i\in\Phi_u}\{L^{-\frac{1}{\alpha}}_i\|U_i\|\}$.
Thus, the distribution of $R_{\star}^{-\frac{1}{\alpha}}$ can reflect how the LoS effect impacts the distribution of the shortest distance between the origin and set $\Phi_u$. By considering $W=1$ in~\eqref{Eqn:CDFR*1}, we can obtain $F^c_{R^{-\frac{2}{\alpha}}_{\star}}(y)$ as  shown in the following:
\begin{align}\label{Eqn:CCDFInvR*3}
	F^c_{R^{-\frac{2}{\alpha}}_{\star}}(y) = \exp\left(-\pi \lambda\omega  y\right),
\end{align} 
i.e., $R^{-\frac{2}{\alpha}}_{\star}\sim\exp(\pi\lambda\omega)$, which reveals the following point set
\begin{align}
	\widetilde{\Phi}_u\defn\left\{\widetilde{U}_i\in\mathbb{R}^2\times\mathbb{R}_+: \widetilde{U}_i=L^{-\frac{1}{\alpha}}_iU_i, L_i\in\{1,\ell\}, U_i\in\Phi_u\right\} \label{Eqn:Equ3DPointProcess}
\end{align}
can be viewed as a thinning PPP from $\Phi_x$ with density $\lambda\omega$. When $\ell=0$, $R^{-\frac{1}{\alpha}}_{\star}$ is the shortest distance of the LoS link from the origin to set $\Phi_u$ and $F^c_{R^{-\frac{2}{\alpha}}_{\star}}(x)$ in~\eqref{Eqn:CCDFInvR*3} reduces to $e^{-\pi\lambda \mathbb{E}[\rho(\Theta)\cos^2(\Theta)]y}$. Therefore,  \textit{in the APIL scenario the LoS points in $\Phi_u$ are equivalent to a 2D homogeneous PPP of density $\lambda \mathbb{E}[\cos^2(\Theta)\rho(\Theta)]$}.

These above observations learned from $R_{\star}$ considerably help us understand some fundamental and intrinsic properties of $\Phi_u$ and they are very useful for the following analyses.

\section{Modeling and Analysis of A 3D UAV-Enabled Cellular Networks Using $\Phi_u$}\label{Sec:UAVNetworkModel}

In this section, we employ the proposed 3D point process $\Phi_u$ in~\eqref{Eqn:3DPP} to model the random locations of UAVs in a cellular network, as shown in Fig.~\ref{Fig:SystemModel}. The salient feature of using $\Phi_u$ to model the 3D locations of the UAVs, as we will see, is not only  to generally characterize the distribution of the UAVs hovering in the sky but also to properly and tractably analyze the performances of a UAV-enabled cellular network. Our focus in this section is on the study of the coverage performance of a UAV-enabled cellular network in which a tier of UAVs are deployed in the sky that serve as \textit{aerial} base stations in the network and the locations of the UAVs are modeled by $\Phi_u$, i.e., $U_i$ in $\Phi_u$ denotes UAV $i$ and its location in the network. Suppose there is a typical user located at the origin and each user in the UAV-enabled cellular network associates with a UAV that provides it with the (averaged) strongest received signal power. Namely, the UAV associated with the typical user is given by
\begin{align}\label{Eqn:AssoUAV}
	U_{\star} &\defn\arg\max_{i:U_i\in\Phi_u} \mathbb{E}\left[PL_iG_i\|U_i\|^{-\alpha}|U_i\right]\nonumber\\ &=\arg\max_{i:U_i\in\Phi_u} \frac{P\mathbb{E}[G]L_i}{\|U_i\|^{\alpha}} =\arg\max_{i:U_i\in\Phi_u} \frac{L_i}{\|U_i\|^{\alpha}},
\end{align} 
where $P$ is the transmit power of each UAV, $G_i\sim\exp(1)$ denotes the fading channel gain between the typical user and $U_i$, $\alpha>2$ denotes the path-loss exponent in this context, and $L_i$, as already defined in~\eqref{Eqn:Weight3DDis}, is used to characterize the LoS and NLoS channel effects in the channel between $U_i$ and the typical user. The second equality in~\eqref{Eqn:AssoUAV} is due to considering the independence between $G_i$ and $U_i$ as well as conditioning on $U_i$, and the third equality is owing to  removing constants $P$ and $\mathbb{E}[G]$ does not affect the result of finding $U_{\star}$.

\subsection{The SINR Model}
Let $I_0$ be the aggregated interference power received by the typical user that does not include the signal power from $U_{\star}$ so that it can be written as
\begin{align}\label{Eqn:Interference}
	I_0 \defn \sum_{i:U_i\in\Phi_u\setminus U_{\star}} P G_i L_i\|U_i\|^{-\alpha}.
\end{align}
All $G_i$'s are assumed to be i.i.d. and they are independent of all $L_i$'s and $U_i$'s. Note that each UAV is associated with at least one user so that the ``void'' UAV phenomenon is not modeled in $I_0$~\cite{CHLLCW1502}\cite{CHLLCW16}. In addition, each UAV allocates different resource blocks (RBs) to different users associating with it, i.e., no users associating with the same UAV can share the same RB. 

Each UAV is assumed to be equipped with $N$ antennas whereas each user is equipped with a single antenna. According to~\eqref{Eqn:AssoUAV} and~\eqref{Eqn:Interference}, if each UAV is able to perform transmit beamforming to its user, the signal-to-interference plus noise power ratio (SINR) of the typical user can be defined as 
\begin{align}\label{Eqn:DefSINR}
	\gamma_0 \defn \frac{PG_{\star}L_{\star}\|U_{\star}\|^{-\alpha}}{I_0+\sigma_0},
\end{align}
where $G_{\star}\sim\text{Gamma}(N,1)$ is the fading channel gain from $U_{\star}$ to the typical user, $L_{\star}\in\{1,\ell\}$ has the same distribution as $L_i$, and $\sigma_0$ denotes the thermal noise power from the environment. The downlink coverage (probability) of a user in the network can thus be defined as
\begin{align}\label{Eqn:KjointCovProb}
	p_{cov}\defn \mathbb{P}\left[\gamma_0 \geq \beta\right]= \mathbb{P}\left[\frac{ PG_{\star}L_{\star}\|U_{\star}\|^{-\alpha}}{I_0+\sigma_0}\geq \beta \right],
\end{align}
where $\beta>0$ is the SINR threshold for successful decoding. In the following, we will analyze $p_{cov}$ by considering whether the elevation angle and the projection of each UAV are independent or not. In the following section, we will employ the model of the UAV-enabled cellular network proposed in this section to analyze the coverage performance of the network.

 \subsection{Downlink Coverage Analysis}\label{SubSec:PerAnaAPIL}

In this subsection, we would like to study the downlink coverage $p_{cov}$ in~\eqref{Eqn:KjointCovProb} by considering the APIL scenario, i.e., elevation angle $\Theta_i$ and projection $X_i$ of UAV $U_i$ are independent for all $i\in\mathbb{N}_+$. The following proposition specifies the analytical result of $p_{cov}$ in this scenario.
\begin{proposition}\label{Prop:CovProbIndep}
	If the APIL scenario is considered, the downlink coverage defined in~\eqref{Eqn:KjointCovProb} can be found as
	\begin{align}\label{Eqn:CovProbIndep}
		p_{cov} =& \frac{\dif^{N-1}}{\dif \tau^{N-1}}\mathbb{E}\bigg[\frac{\tau^{N-1}}{(N-1)!}\exp\bigg(-\frac{\sigma_0 D_{\star}^{\frac{\alpha}{2}}}{\tau P}\nonumber\\
		&-\pi\lambda\omega D_{\star}\mathcal{I}_G\left(\frac{1}{\tau},\frac{2}{\alpha}\right)\bigg)\bigg]\bigg|_{\tau=\frac{1}{\beta}}, 
	\end{align} 
	where $D_{\star}\sim\exp(\pi\lambda\omega)$ and function $\mathcal{I}_G(u,v)$ is defined as
	\begin{align}
		\mathcal{I}_G\left(u,v\right)\defn u^{v}\left(\frac{\pi v}{\sin(\pi v)}-\int_0^{u^{-v}}\frac{\dif r}{1+r^{\frac{1}{v}}}\right). \label{Eqn:IfunNoW}
	\end{align}
\end{proposition}
\begin{IEEEproof}
Please refer to Appendix D in~\cite{CHLDCL21} for the complete proof, which is omitted due to limited space. 
\end{IEEEproof}
\noindent We adopt an exponential RV $D_{\star}$ with mean $1/\pi\lambda\omega$ in~\eqref{Eqn:CovProbIndep} to make $p_{cov}$ show in a neat form so as to clearly see how $p_{cov}$ is impacted by $D_{\star}$ and other network parameters. The physical meaning of $D_{\star}$ is the square of the shortest distance between the typical user and set $\widetilde{\Phi}_u$, i.e., $D_{\star}\defn\|\widetilde{U}_{\star}\|^2\stackrel{d}{=}L^{-\frac{2}{\alpha}}_{\star}\|U_{\star}\|^2$ where $\widetilde{U}_{\star}$ is the nearest point in $\widetilde{\Phi}_u$ to the typical user and $\stackrel{d}{=}$ stands for the equivalence in distribution. In other words, $p_{cov}$ is highly dependable upon the distribution of elevation angle $\Theta$ and $\ell$ for a given density $\lambda$ because the distribution of $D_{\star}$ is parameterized with $\lambda\omega$. To make this point much clear, we use Jensen's inequality to find a lower bound on $p_{cov}$ in~~\eqref{Eqn:CovProbIndep} as
\begin{align}\label{Eqn:CovProbIndepIneq1}
	p_{cov}\geq & \frac{1}{(N-1)!}\frac{\dif^{N-1}}{\dif \tau^{N-1}}\bigg\{\tau^{N-1}\exp\bigg[-\frac{N\sigma_0\Gamma\left(1+\frac{\alpha}{2}\right)}{\tau P(\pi\lambda\omega)^{\frac{\alpha}{2}}}\nonumber\\
	&-\mathcal{I}_G\left(\frac{N}{\tau},\frac{2}{\alpha}\right)\bigg]\bigg\}\bigg|_{\tau=\frac{1}{\beta}},
\end{align}
which reduces to the following neat inequality for $N=1$:
\begin{align}\label{Eqn:CovProbIndepIneq2}
	p_{cov}\geq \exp\left[-\frac{\beta\sigma_0\Gamma\left(1+\frac{\alpha}{2}\right)}{ P(\pi\lambda\omega)^{\frac{\alpha}{2}}}-\mathcal{I}_G\left(\beta,\frac{2}{\alpha}\right)\right].
\end{align}
The inequalities in~\eqref{Eqn:CovProbIndepIneq1} and \eqref{Eqn:CovProbIndepIneq2}  apparently show that increasing $\lambda\omega$ improves $p_{cov}$. This is because users are able to associate with a nearer UAV and receive stronger power from the UAV when deploying UAVs more densely even though more interference is generated as well. Also, $p_{cov}$ improves whenever  $\lambda\omega$ can be maximized by optimizing the distribution of $\Theta$. We will demonstrate some numerical results in Section~\ref{Sec:Simulation} to show how $p_{cov}$ varies with different distribution cases of $\Theta$.

An effective method to significantly improve the coverage of users is to make users associate with multiple UAVs so that the UAVs can do coordinated multi-point (CoMP) joint transmission. The upper limit of the downlink coverage of a user associating with multiple UAVs can be achieved when all the UAVs are coordinated to jointly transmit to the user at the same time, which is referred as to the \textit{cell-free downlink coverage}. Since perfectly coordinating and synchronizing all the UAVs in a large-scale network to do \textit{coherent} transmission is hardly possible in practice, \textit{non-coherent joint transmission} is a feasible way for all the UAVs to jointly achieve the cell-free downlink coverage in that it has lower implementation complexity and does not require high backhaul capacity if compared with its coherent counterpart. When all the UAVs perform non-coherent CoMP joint transmission to a user, the cell-free downlink coverage of the user can be defined as~\cite{DLHSBCEH12,RTSSJGA14}
\begin{align}\label{Eqn:DefnCellFreeCovProb}
	p^{cf}_{cov} \defn \mathbb{P}\left[\frac{P\sum_{i:U_i\in\Phi_u}G_iL_i\|U_i\|^{-\alpha}}{\sigma_0} \geq \beta \right],
\end{align}
where $G_i\sim\text{Gamma}(N,1)$ for all $i\in\mathbb{N}_+$ since all the UAVs can do transmit beamforming to the user. The explicit result of $p^{cf}_{cov}$ can be found as shown in the following proposition.
\begin{proposition}\label{Prop:CellFreeCovProbAPIL}
	If all the UAVs are deployed based on the APIL scenario and coordinated to do non-coherence joint transmission, the cell-free downlink coverage defined in~\eqref{Eqn:DefnCellFreeCovProb} is derived as
	\begin{align}\label{Eqn:CellFreeCov1}
		p^{cf}_{cov}= 1- \mathcal{L}^{-1}\bigg\{ & \frac{1}{s}\exp\bigg[-\frac{\pi\lambda s^{\frac{2}{\alpha}}\omega}{(N-1)!}\Gamma\left(N+\frac{2}{\alpha}\right)\nonumber\\
		&\Gamma\left(1-\frac{2}{\alpha}\right)\bigg]\bigg\}\left(\frac{\beta\sigma_0}{P}\right),
	\end{align}
	which reduces to the following closed-form result for $\alpha=4$:
	\begin{align}\label{Eqn:CellFreeCov2}
		p^{cf}_{cov} = \mathrm{erf}\left(\frac{\pi^{\frac{3}{2}}\lambda\omega}{2(N-1)!}\sqrt{\frac{P}{\beta\sigma_0}}\Gamma\left(N+\frac{1}{2}\right)\right),
	\end{align}
	where $\mathrm{erf}(z)\defn \frac{2}{\sqrt{\pi}}\int_0^z e^{-t^2}\dif t$ is the error function for $z>0$.
\end{proposition}
\begin{IEEEproof}
The proof is omitted here due to limited space.
\end{IEEEproof}

\section{Numerical Results}\label{Sec:Simulation}

\begin{table}[!t] 
	\centering
	\caption{Network Parameters for Simulation~\cite{AHKSSL14}}\label{Tab:SimPara}
	\begin{tabular}{|c|c|}
		\hline Transmit Power  (mW)  $P$  & $50$ \\ 
		\hline Density of set $\Phi_x$ (points (UAVs)/m$^2$) $\lambda_x$  & $1.0\times 10^{-7}\sim 1.0\times 10^{-5}$   \\ 
		\hline Number of Antennas $N$ & $1$, $4$, $8$, $\infty$ (or see figures) \\ 
		\hline Noise Power (dBm) $\sigma_0$ & $-92.5$ \\ 
		\hline Path-loss Exponent $\alpha$ & $2.75$\\ 
		\hline Parameters $(c_1,c_2)$ in \eqref{Eqn:LoSProb} for Suburban & $(24.5811,39.5971)$ \\
		\hline NLoS Channel Attenuation Factor $\ell$ & $0.25$ \\
		\hline SINR Threshold (dB) $\beta$  & $-10$ (or see figures) \\ 
		\hline
	\end{tabular}
		\vspace{-10pt} 
\end{table}

\begin{figure}[t!]
	\centering
	\includegraphics[width=0.5\textwidth, height=2in]{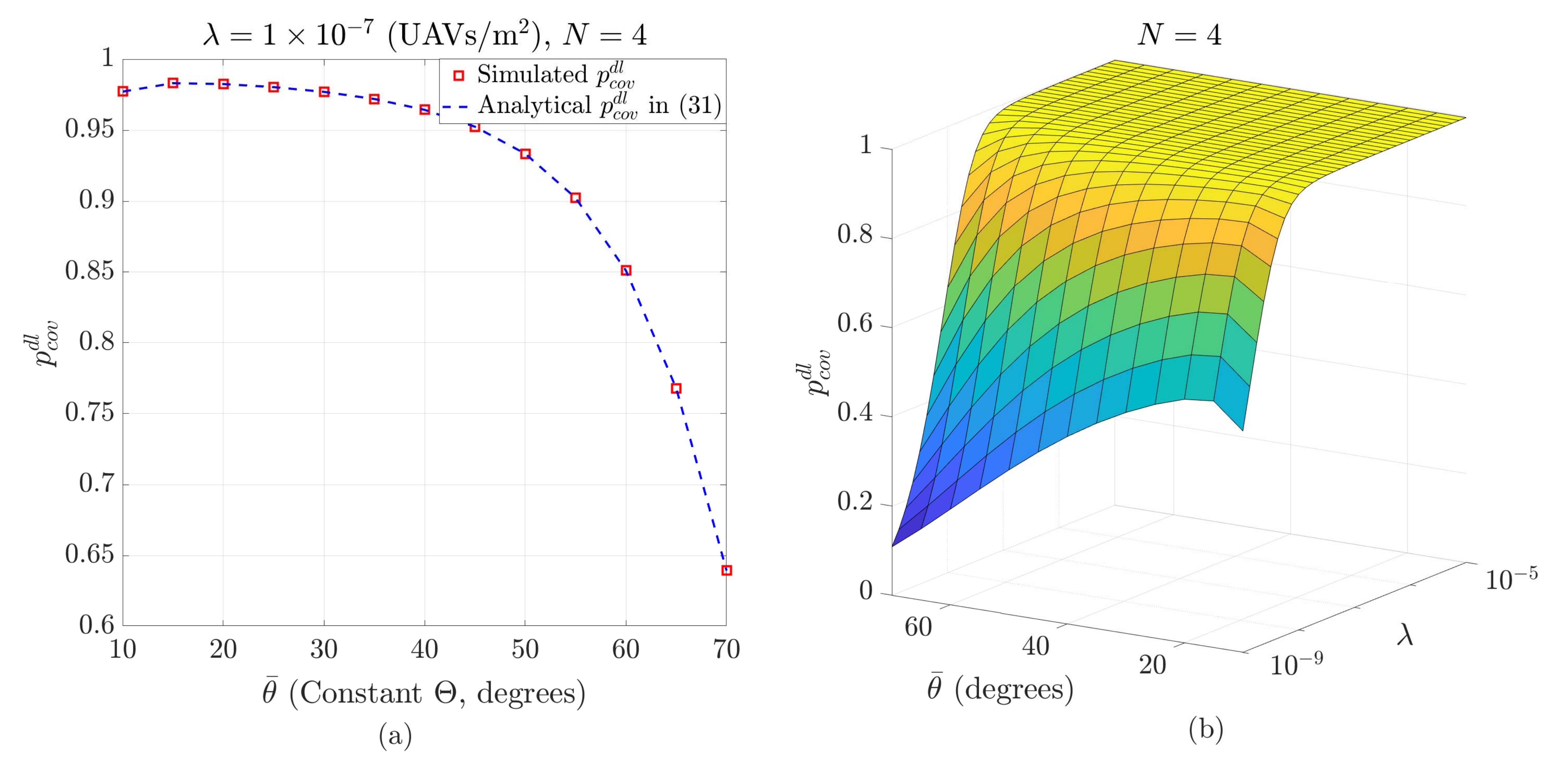}
	\caption{Simulation results of $p^{dl}_{cov}$ for the APIL scenario when $N=4$ and the elevation angle  of each UAV is a constant $\overline{\theta}$ with respect to the origin (i.e., $\tan(\Theta)\sim\text{Gamma}(a,a/\tan(\overline{\theta}))$ as $a\rightarrow\infty$): (a) 2D simulation results of $p^{dl}_{cov}$ versus elevation angle $\overline{\theta}$ for $\lambda=1\times 10^{-7}$ (UAV /m$^2$), (b) 3D simulation results of $p^{dl}_{cov}$ versus density $\lambda$ and elevation angel $\overline{\theta}$.}
	\vspace{-10pt}
	\label{Fig:APIL_ConstAng}
\end{figure}

\begin{figure}[t!]
	\centering
	\includegraphics[width=0.5\textwidth, height=2in]{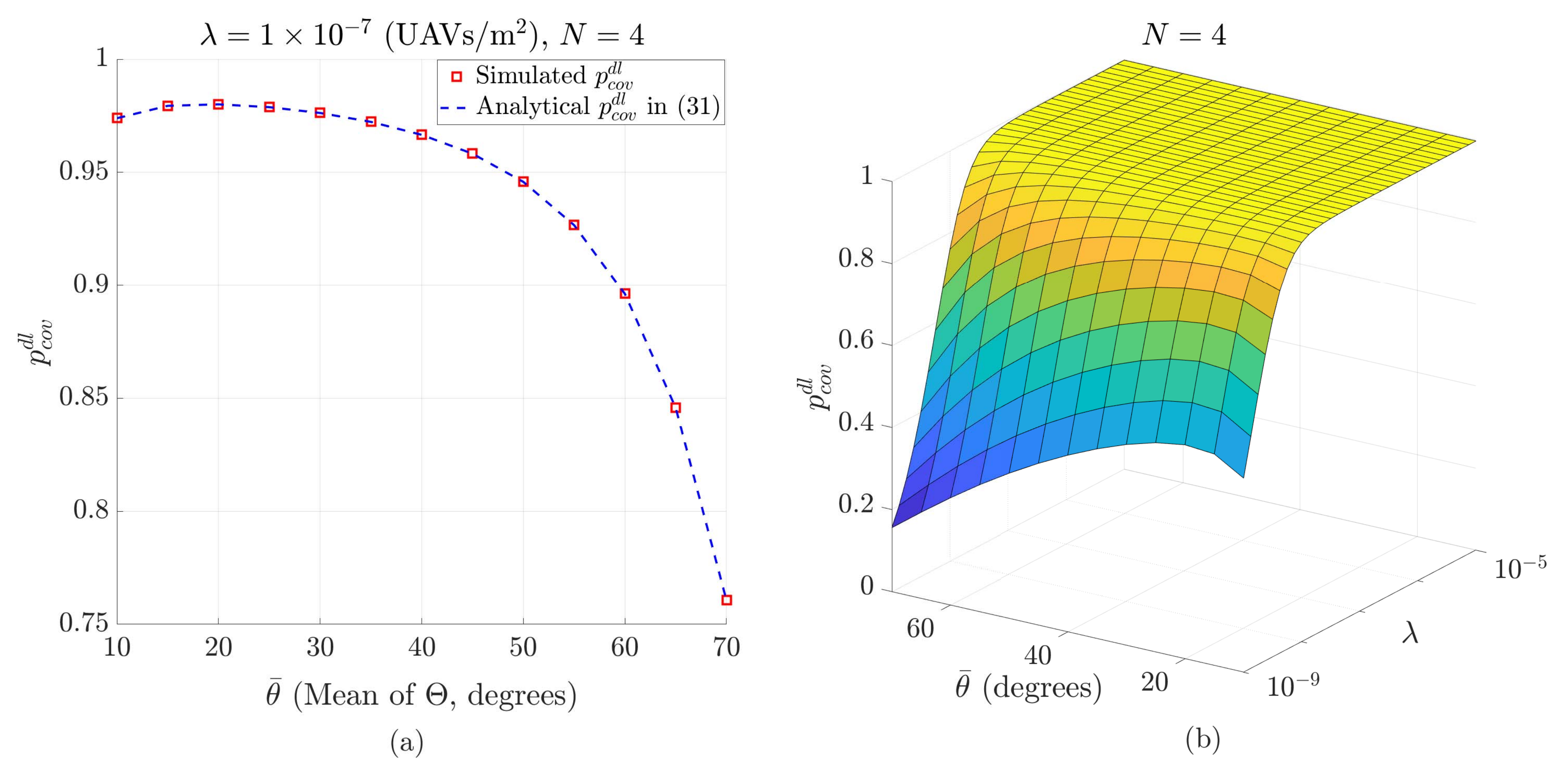}
	\caption{Simulation results of $p^{dl}_{cov}$ for the APIL scenario when $N=4$ and the elevation angle of each UAV is a Gamma RV with shape parameter $a$ and rate parameter $a/\tan(\overline{\theta})$, i.e., $\tan(\Theta)\sim\text{Gamma}(a,a/\tan(\overline{\theta}))$: (a) 2D simulation results of $p^{dl}_{cov}$ versus elevation angle $\Theta$ for $\lambda=1\times 10^{-7}$ (UAV/m$^2$), (b) 3D simulation results of $p^{dl}_{cov}$ versus density $\lambda$ and mean of elevation angel $\overline{\theta}$.}
	\vspace{-10pt}
	\label{Fig:APIL_GammaAng}
\end{figure}

\begin{figure}[t!]
	\centering
	\includegraphics[width=0.48\textwidth, height=2.2in]{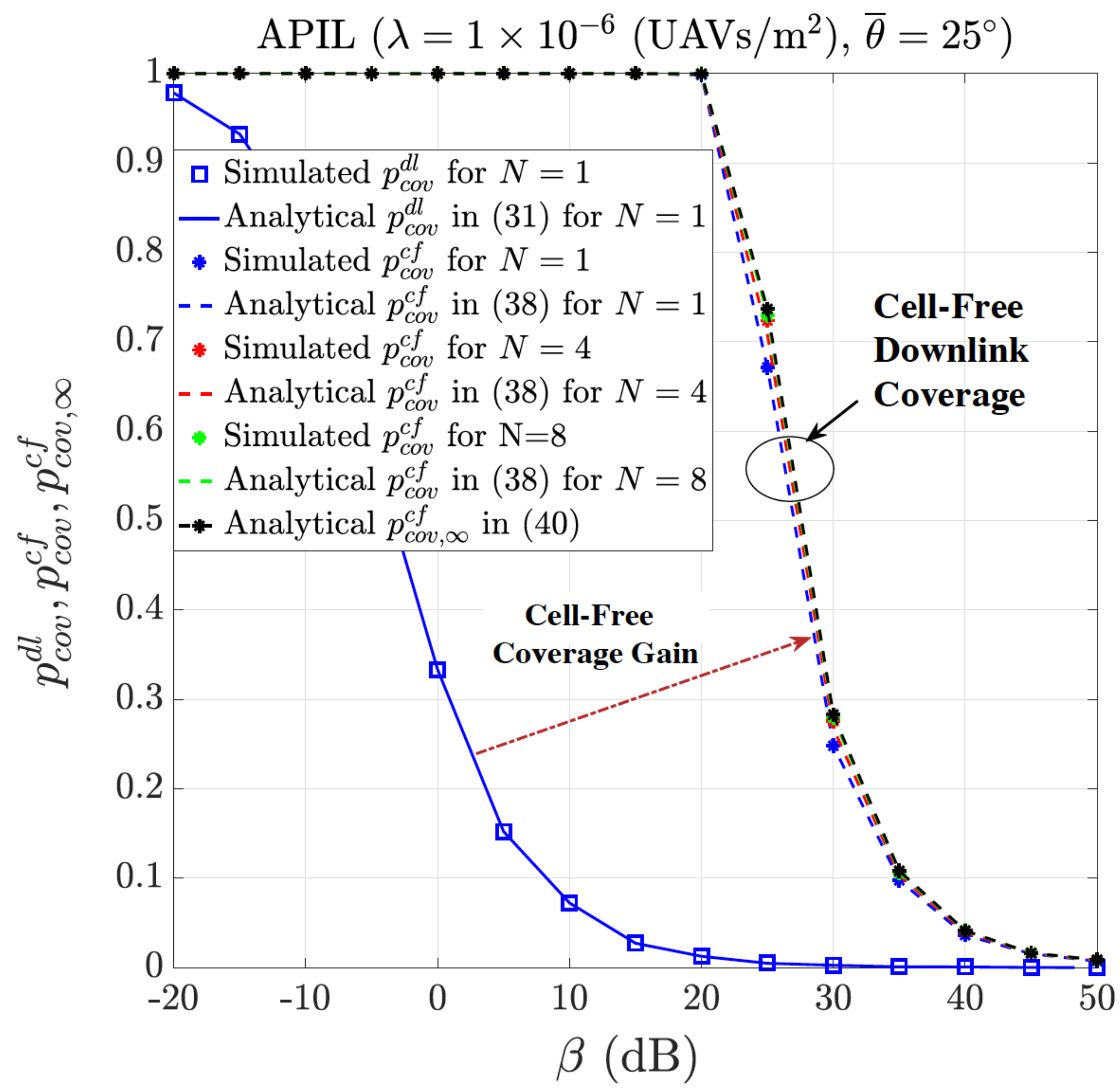}
	\caption{Simulation results of cell-free downlink coverage $p^{cf}_{cov}$ for $\lambda=1\times 10^{-6}$ (UAV/m$^2$) and $N=1, 2, 4, 8, \infty$: Simulation results of $p^{cf}_{cov}$ versus SINR threshold $\beta$ for the APIL scenario and the elevation angle of each UAV is a constant equal to $\overline{\theta}=25^{\circ}$.}
	\label{Fig:CellFreeCovProb}
		\vspace{-15pt}
\end{figure}

In this section, we will provide some numerical results to verify the previous analytical results of the coverage for the APIL scenario in Figs.~\ref{Fig:APIL_ConstAng} and~\ref{Fig:APIL_GammaAng}. The numerical results of the downlink cell-free coverage will be presented in Fig.~\ref{Fig:CellFreeCovProb}. The network parameters adopted for simulation are shown in Table~\ref{Tab:SimPara}. We consider the tangent of the elevation angle of a UAV is a Gamma RV with shape parameter $a$ and rate parameter $b$ (i.e., $\tan(\Theta)\sim\text{Gamma}(a,b)$) because using such a Gamma RV to model $\tan(\Theta)$ is able to generally characterize different distributions by setting different values of $a$ and $b$ so that appropriately adjusting $a$ and $b$ can make $\Theta$ reasonably distribute between $0$ and $\frac{\pi}{2}$. For example, $\tan(\Theta)$ becomes deterministic and equal to $\tan(\overline{\theta})$ such that $\Theta$ is equal to constant $\overline{\theta}$ if $b=a/\tan(\overline{\theta})$ and $a\rightarrow\infty$ and it becomes an exponential RV with rate parameter $1/b$ if $a=1$. Figs.~\ref{Fig:APIL_ConstAng} and~\ref{Fig:APIL_GammaAng} show the simulation results of the downlink coverage $p^{dl}_{cov}$ when $\tan(\Theta)$ is a constant and a Gamma RV, respectively. As we can see, the simulation results of $p^{dl}_{cov}$ in Figs.~\ref{Fig:APIL_ConstAng}(a) and~\ref{Fig:APIL_GammaAng}(a) do not differ much when $\overline{\theta}<45^{\circ}$, which reveals that \textit{in general $p^{dl}_{cov}$ is insensitive to the distribution of $\Theta$} when the mean of $\Theta$ is not very large. In fact, this phenomenon can be inferred from~\eqref{Eqn:CovProbIndep} in that $p^{dl}_{cov}$ is affected by the distribution of $\Theta$ through $\omega$ in~\eqref{Eqn:Density3DPPInd} that is insensitive to the distribution of $\Theta$ when the mean of $\Theta$ is not large. Realizing this phenomenon is quite useful since we can quickly and accurately calculate $p^{dl}_{cov}$ using the mean of the elevation angle of UAVs in~\eqref{Eqn:CovProbIndep} without knowing the real distribution of $\Theta$, which is in general not easy to find in practice. 

Figs.~\ref{Fig:APIL_ConstAng}(a) and~\ref{Fig:APIL_GammaAng}(a) validate the correctness and accuracy of the expression in~\eqref{Eqn:CovProbIndep} since the curve of the analytical result of $p^{dl}_{cov}$ in~\eqref{Eqn:CovProbIndep} completely coincides with the curve of the simulated result of $p^{dl}_{cov}$. Moreover, there exists an optimal value of the mean of $\Theta$ about $20^{\circ}$ for $\lambda=1\times 10^{-7}$ (UAVs/m$^2$), which maximizes $p^{dl}_{cov}$. Note that $p^{dl}_{cov}$ decreases as the mean of $\Theta$ increases over $20^{\circ}$ since the downlink SINR is now dominated by the interference in this situation even though the received signal power also increases. The 3D plots in Figs~\ref{Fig:APIL_ConstAng}(b) and Fig.~\ref{Fig:APIL_GammaAng}(b) further show how $p^{dl}_{cov}$ varies with the mean of $\Theta$ and $\lambda$. Generally speaking, the optimal value of the mean of $\Theta$ that maximizes $p^{dl}_{cov}$ changes with density $\lambda$ and $p^{dl}_{cov}$ converges up to a constant as $\lambda$ goes to infinity, i.e., $p^{dl}_{cov}$ barely depends on $\lambda$ as the network is dense and interference-limited.

According to Fig.~\ref{Fig:CellFreeCovProb} that shows the numerical results of $p^{cf}_{cov}$, we can observe a few interesting and important phenomena.  First, the analytical results of the downlink cell-free coverages in the figure perfectly coincide with their corresponding simulated results, which validates the correctness of the expression in~\eqref{Eqn:CellFreeCov1}.  Second, the downlink cell-free coverages for different numbers of antennas are almost identical and this reveals that UAVs do not need to install multiple antennas to improve their coverage in the cell-free scenario so that UAVs can become lighter so as to save more power when flying. 
Third, the downlink cell-free coverage $p^{cf}_{cov}$ significantly outperforms the downlink coverage $p^{dl}_{cov}$, as can be seen in the figure. 

\section{Conclusion}
In the past decade, using 2D PPPs to model large-scale cellular networks had given rise to a great success in tractably analyzing the generic performance metrics of cellular networks. Nevertheless, straightforwardly employing a 3D PPP to deploy UAVs in a cellular network not only poses an unrealistic constraint on the path-loss exponent of 3D path-loss channel models, but also ignores a spatial deployment limitation in a cellular network. To tackle this issue, this paper proposes a 3D point process whose projections consist of a 2D homogeneous PPP and altitudes are the marks of the 2D homogeneous PPP. The fundamental properties of the proposed 3D point process are studied for the APIL scenario and they pave a tractable way to analyze the downlink coverage of a UAV-enabled cellular network modeled by the proposed 3D point process. The downlink coverage is explicitly derived and its closed-form expression is also found for a special channel condition. In addition, cell-free downlink coverage and its upper limits are also derived when all the UAVs in the network can do non-coherence joint transmission. 

\appendix
\numberwithin{equation}{section}
\setcounter{equation}{0}

Consider the APIL scenario so that $X_i$ and $\Theta_j$ are independent for all $i,j\in\mathbb{N}_+$. Since $\|U_i\|=\|X_i\|\sec(\Theta_i)$, the CDF of $R_{\star}$ defined in~\eqref{Eqn:Weight3DDis} can be written as
\begin{align}
	& F_{R_{\star}}(r)= \mathbb{P}\left[\max_{i:U_i\in\Phi_u} \left\{\frac{W_iL_i}{[\|X_i\|\sec(\Theta_i)]^{\alpha}}\right\}\leq r\right]\\
	&\stackrel{(a)}{=}\mathbb{E}\left\{\prod_{i:U_i\in\Phi_u}\mathbb{P}\left[\frac{W_iL_i}{[\|X_i\|\sec(\Theta_i)]^{\alpha}}\leq r\right]\right\}\nonumber\\
	&\stackrel{(b)}{=}\exp\left(-2\pi\lambda\int_0^{\infty}\mathbb{P}\left[\frac{WL}{[x\sec(\Theta)]^{\alpha}}\geq r\right]x\dif x\right), \label{Eqn:AppCDFR*1}
\end{align}
where $(a)$ follows from the fact that all $W_iL_i[\|X_i\|\sec(\Theta_i)]^{-\alpha}$'s are independent and $(b)$ is obtained by first considering the independence between all RVs $W_i$, $L_i$, $\|X_i\|$, and $\Theta_i$ for all $i\in\mathbb{N}_+$ and then applying the probability generation functional (PGFL) of a homogeneous PPP to $\Phi_x$\footnote{Note that the subscript $i$ in $(a)$ is dropped in $(b)$ for notation simplification and such a subscript dropping is used throughout this paper whenever there is no notation ambiguity.}.  According to \eqref{Eqn:LoSProb}, $\mathbb{P}[WL[x\sec(\Theta)]^{-\alpha}\geq r|\Theta]$ can be further expressed as 
\begin{align*}
	&\mathbb{P}\left[\frac{WL}{[x\sec(\Theta)]^{\alpha}}\geq r \bigg|\Theta\right] = \mathbb{P}\left[\left(\frac{W}{r}\right)^{\frac{1}{\alpha}}\cos(\Theta)\geq x\bigg|\Theta\right]\rho\left(\Theta\right)\\
	&+\mathbb{P}\left[\left(\frac{\ell W}{r}\right)^{\frac{1}{\alpha}}\cos(\Theta)\geq x\bigg|\Theta\right] \left[1-\rho\left(\Theta\right)\right]. 
\end{align*}
Therefore, we can have the following:
\begin{align*}
	&2\int_0^{\infty}\mathbb{P}\left[\frac{WL}{[x\sec(\Theta)]^{\alpha}}\geq r\bigg|\Theta\right]x\dif x \\
	&  = \rho(\Theta) \int_0^{\infty} \mathbb{P}\left[\left(\frac{W}{r}\right)^{\frac{1}{\alpha}}\cos(\Theta)\geq x\bigg|\Theta\right]\dif x^2+[1-\rho(\Theta)]\\
	& \times \int_0^{\infty} \mathbb{P}\left[\left(\frac{\ell W}{r}\right)^{\frac{1}{\alpha}}\cos(\Theta)\geq x\bigg|\Theta\right]\dif x^2  \\
	& = \cos^2(\Theta)\left[\rho(\Theta) +[1-\rho(\Theta)]\ell^{\frac{2}{\alpha}}\right]\mathbb{E}\left[W^{\frac{2}{\alpha}}\right]r^{-\frac{2}{\alpha}}
\end{align*}
since $\int_0^{\infty}\mathbb{P}[Z\geq z]\dif z=\mathbb{E}[Z]$ for a non-negative RV $Z$. This gives rise to the following result:
\begin{align*}
	& 2\int_0^{\infty}\mathbb{P}\left[\frac{WL}{[x\sec(\Theta)]^{\alpha}}\geq r\right]x\dif x \\
	&=\mathbb{E}\left\{\cos^2(\Theta)\left[\rho(\Theta) \left(1-\ell^{\frac{2}{\alpha}}\right)+\ell^{\frac{2}{\alpha}}\right]\right\}\mathbb{E}\left[W^{\frac{2}{\alpha}}\right] r^{-\frac{2}{\alpha}},
\end{align*}
and then substituting this identity into~\eqref{Eqn:AppCDFR*1} yields the expression in~\eqref{Eqn:CDFR*1}.


\bibliographystyle{IEEEtran}
\bibliography{IEEEabrv,Ref_3DUAVCellular} 

\end{document}